\documentstyle[prl,aps,multicol,epsfig]{revtex}

\begin{document}

\title{  Anomalous roughening of Hele-Shaw flows with quenched disorder }

\author{J. Soriano$^1$, J.J. Ramasco$^{2,3}$, M.A.  Rodr\'{\i}guez$^2$,
A. Hern\'{a}ndez-Machado$^{1,4}$,  and J. Ort\'{\i}n$^1$\\}

\address{  $^1$ Departament ECM, Facultat de F\'{\i}sica, Universitat de
Barcelona\\ Diagonal  647, E-08028 Barcelona, Spain \\ $^2$ Instituto de
F\'{\i}sica de Cantabria, CSIC\\  Avenida Los Castros, E-39005 Santander,
Spain.\\ $^3$ Departamento de F\'{\i}sica Moderna, Universidad de Cantabria,
E-39005 Santander, Spain. \\$^4$ Groupe de Physique des Solides,
Universit\'{e} Pierre-et-Marie-Curie\\ Tour 23, 2 place Jussieu, 75251 Paris
Cedex 05,  France.}

\date{\today}

\maketitle

\begin{abstract}

The kinetic roughening of a stable oil--air interface, moving in a Hele--Shaw
cell which contains a quenched columnar disorder (tracks) has been studied. A
capillary effect is responsible for the dynamic evolution of the resulting
rough interface, which exhibits anomalous scaling. The three independent
exponents needed to characterize the anomalous scaling are determined
experimentally. The anomalous scaling is explained in terms of the initial
acceleration and subsequent deceleration of the interface tips in the tracks
coupled by mass conservation. A phenomenological model that reproduces the
measured global and local exponents has been introduced.

\end{abstract}






\begin{multicols}{2}

\narrowtext

The kinetic roughening of surfaces and interfaces in nonequilibrium
conditions has received a great deal of attention in last years.  This
is a subject of technological importance in surface growth,
and of fundamental relevance as an example of scaling far from
equilibrium \cite{meakin-kardar-krug}.

Roughening originates from the competition between stabilizing and
destabilizing mechanisms at different length scales.  As a result of the
competition, the correlation length of the interfacial fluctuations
along an initially flat front increases monotonously with time, until it
reaches the system size and the roughness saturates.  The statistical
properties of the roughening front are usually given in terms of the
({\it rms}) width $w$ of the interfacial fluctuations.  In the absence
of characteristic scales, many systems follow the dynamic scaling
hypothesis of Family--Vicsek (FV) \cite{Family}, in which short and long
length scales have the same scaling behavior:  $w(l) \sim t^{\beta}$ for
$t<t_{\times}$, $w(l) \sim l^\alpha$ for $t>t_{\times}$ and $t_{\times}
\sim l^z$.  Here $l$ is the lateral size of the window used to measure
$w$, $t_{\times}$ is the saturation time, $\beta$ the growth exponent,
$\alpha$ the roughness exponent, and $z$ the dynamic exponent.  The
scaling relation $\alpha = z \beta$ reduces to two the number of
independent exponents that characterize the universality class of the
problem under study.

In recent years, however, experiments on molecular beam epitaxy
\cite{MBE-Yang}, sputtering \cite{Sputtering-Jeffries}, fracture
mechanics \cite{fracture}, and
electrodeposition \cite{ED-Huo} have revealed a rich variety of
situations in which the scaling of the global and the local interface
fluctuations are substantially different.  Long length scales are then
characterized by the same FV scaling exponents, while short length
scales follow the so--called {\it anomalous scaling ansatz}
\cite{miguel}:  $w(l,t) \sim t^{\beta}$ for $t<t_l$, $w(l,t) \sim
l^{\alpha_{loc}}t^{\beta^{\ast}}$ for $t_l < t < t_{\times}$, and
$w(l,t) \sim l^{\alpha_{loc}}L^{\alpha-\alpha_{loc}}$ for $t_{\times} <
t$.  The local time $t_l \sim l^{z}$, and the saturation time
$t_{\times} \sim L^{z}$, where $L$ is the system size in the direction
along the front.  Here $\beta^{\ast}$ and $\alpha_{loc}$ are new {\it
local} growth and roughness exponents, respectively, which characterize
the scaling at short length scales ($l \ll L)$.  There is a new scaling
relation $\alpha - \alpha_{loc} = z \beta^{\ast}$.  Since now there are five
exponents and two scaling relations, the anomalous scaling is
characterized by three independent exponents.  Usually the roughness
exponents are determined through the power spectrum, which is less
sensitive than $w$ to finite size effects.  The spectrum scales as
$S(k,t) = k^{-(2\alpha + 1)} s_{A}(k t^{1/z}) $, where $s_{A}$ obeys
$s_{A}(u) \sim u^{2 \alpha +1}$ when $u \ll 1$ and $s_{A}(u) \sim
u^{2\theta}$ when $u \gg 1$.  Here $\theta=\alpha - \alpha_{loc}$.  The
anomalous scaling leads to the usual FV scaling when $\theta = 0$ (short
and long length scales are characterized by the same roughening
exponents) and $\beta^{\ast} = 0$ (the temporal correlations of the
short length scales disappear).  Recently it has been shown that the
anomalous scaling can be related to the presence of high slopes in the
front that have nontrivial dynamics \cite{juanma}.

\begin{figure}
\begin{center}
\epsfig{file=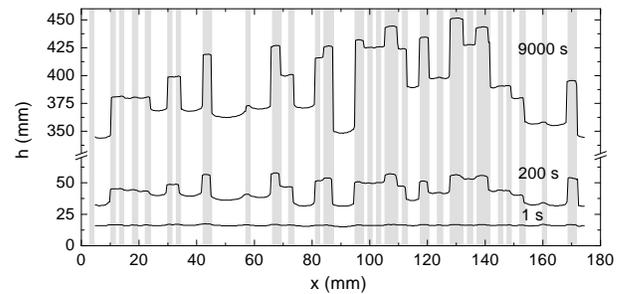,width=8cm}
\caption{Sequence of typical oil--air interfaces. The silicone
oil moves upwards in the picture, and the disorder pattern is represented in
grey. The experimental parameters are  $b=0.36$ mm and $v=0.04$ mm/s.}
\label{Fig:interf}
\end{center}
\end{figure}
In this Letter we report the first observations of anomalous roughening
in experiments of Hele--Shaw flows with quenched disorder.  We consider
an initially flat front between oil and air, driven at a small constant
average velocity $v$ in a horizontal Hele--Shaw cell with a columnar
quenched disorder.  The disorder consists of continuous copper tracks on
a fiberglass substrate in the advancing direction of growth and
randomly distributed in the perpendicular direction (Fig.\
\ref{Fig:interf}).  In this situation, the correlation of the disorder
in the advancing direction is infinite.  The local motion relative to
the average position of the front is driven by capillary forces, caused
by the different curvatures of the advancing front in the third
dimension, depending on whether the oil is on a copper track or on the
fiberglass substrate.  An analysis of the {\it rms} width of the rough
interface at short and long length scales plus a collapse of the power
spectra of the interfacial fluctuations enables an independent
determination of the three exponents characteristic of anomalous
scaling.

We associate the anomalous roughening with the initial acceleration and
subsequent deceleration of the liquid on the copper tracks, caused by
capillarity, and the coupling of the motion over tracks and over
fiberglass due to mass conservation. The anomaly decreases as $v$
increases, and disappears when $v$ is large
enough to override the local acceleration and deceleration
\cite{anomaly}. To explain the experimental results, we propose a
phenomenological model which gives global and local exponents in good
agreement with the experimental exponents.

The experiments have been performed in a horizontal Hele--Shaw cell, $190$ x
$550$ ($L$ x $H$) mm$^{2}$, made of two glass plates $20$ mm thick.
A fiberglass substrate with copper tracks, manufactured using printed
circuit technology, is attached to the bottom plate. The unit copper
track is $d=0.06 \pm 0.01$ mm high
and $1.5 \pm 0.04$ mm wide. The tracks are placed
randomly in a grid of 1.5 mm lattice spacing, along the $x$
direction, up to a 35\% occupation. There are no lateral correlations of
the disorder at distances larger than the width of the unit track.
The distance between the top plate and the
substrate defines the gap spacing $b$, which has been varied in the range
$0.16 \leq b \leq 0.75$ mm. We have used 4 different disorder configurations
and carried out 2 identical runs per disorder configuration. A silicone oil
(kinematic viscosity $\nu=50$ mm$^2$/s, density $\rho=998$ kg/m$^{3}$, and
surface tension oil--air $\sigma=20.7$ mN/m at room temperature) is injected
into one side of the cell at constant volumetric injection rate. The oil
wets similarly and almost completely the surfaces which it is in contact with
(glass, substrate, and copper). The evolution of the interface is
monitored using two CCD cameras mounted along the growth direction and fixed.
The final resolution is $0.37 \times 0.37$ mm$^2$/pixel, providing 4
pixels per unit track in the $x$ direction. We have carried out a series
of reference experiments to calibrate side walls effects and
possible inhomogeneities in gap spacing. Our results make us confident
that these disturbances do not have a relevant effect on the dynamic
scaling presented in this work. Extensive details of the experimental
setup will be given in a forthcoming publication \cite{Soriano-2002-I}.

\begin{figure}
\begin{center}
\epsfig{file=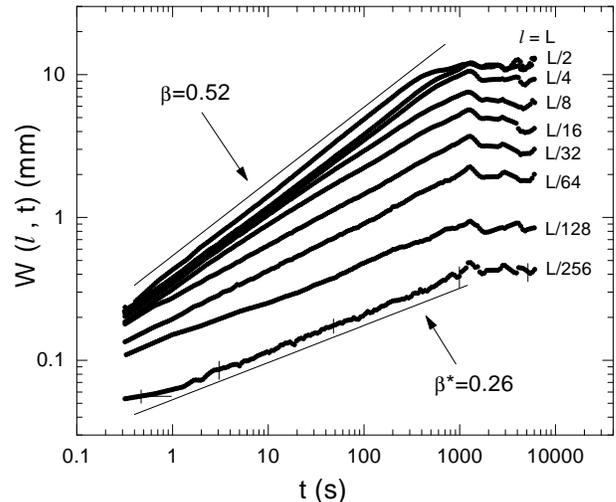,width=8cm}
\caption{Experimental
determination of $\beta$ and $\beta^{\ast}$. The window size $l$
decreases from $L$ to $L/256$. The experimental parameters are
$b=0.36$ mm and $v=0.08$ mm/s.} \label{Fig:beta}
\end{center}
\end{figure}
In Fig.\ \ref{Fig:beta} we present the results for the interfacial width
as a function of time at different length scales $l$, from
$L$ to $L/256$.  We plot $W(l,t)=\left[ w^2(l, t)-w^2(l, 0)
\right]^{1/2}$ to minimize the influence of the initial condition
\cite{Sputtering-Jeffries,subtracted-width}. We obtain clear
power law dependencies, providing $\beta=0.52 \pm 0.02$ at the longest
length scales
and $\beta^{\ast}=0.26 \pm 0.03$ at the two shortest length scales.
It is worth mentioning that a different method,
based on the growth of the local slopes \cite{juanma},
gives the same value of $\beta^{\star}$. The fact that this local
exponent is significantly different from $0$ is a sign of anomalous
scaling of the measured fronts.

\begin{figure}
\begin{center}
\epsfig{file=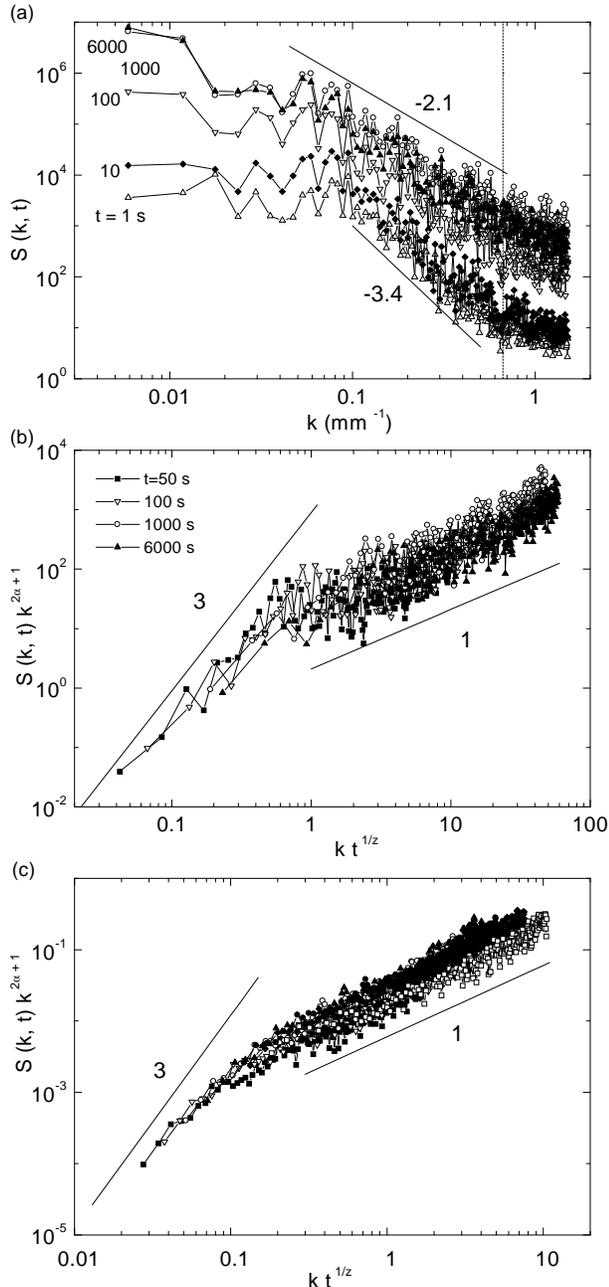,width=8cm}
\caption{(a) Temporal
evolution of the power spectrum for the same experiments as Fig.\
\ref{Fig:beta}. The vertical line gives the value of $k$ associated with the
lateral size of the unit copper track. (b) Collapse of the experimental power
spectra for $t>10$ s. (c) Collapse of the power spectra obtained from numerical
simulations of the interfacial equation (\ref{model}).}
\label{Fig:collapse}
\end{center}
\end{figure}

The temporal evolution of the power spectrum is presented in Fig.\
\ref{Fig:collapse}(a).  At very short times we observe a regime with
$S(k)\sim k^{-3.4}$.  It is a super--rough transient regime created when
the fluid on the tracks reaches its maximum velocity after coming into
contact with the disorder.  The main regime appears further on, with
$S(k)\sim k^{-2.1}$, and a vertical shift that progressively decreases
with time and disappears at saturation ($t \gtrsim 450$ s).  The
vertical shift of the power spectra is a sign of intrinsic anomalous
scaling, in which the spectral exponent \cite{generic} can be identified
with the local roughness exponent, so that $\alpha_{loc}= 0.55 \pm
0.10$.

Using the above procedure, the three independent exponents $\beta$,
$\beta^{\ast}$, $\alpha_{loc}$ have been obtained directly from the
experiments. The other two exponents $\alpha$, $z$ can be obtained from the
scaling relations. In order to verify the whole scaling, however, we prefer to
determine the critical exponents by performing the best collapse of the
spectra compatible with the experimental results. The collapse, presented in
Fig.\ \ref{Fig:collapse}(b), leads to the following set of scaling exponents:
\begin{eqnarray}\label{results}
\nonumber & & \beta=0.50 \pm 0.04,\; \beta^{\ast}=0.25 \pm 0.03,\\
          & & \alpha=1.0 \pm 0.1,\;\alpha_{loc}=0.5 \pm 0.1,\; z=2.0 \pm 0.2.
\end{eqnarray}
We have also looked into the possibility of multiscaling \cite{krug}
through the scaling of the generalized correlations of order $q$, of
the form $\lbrace {(h(x+l,t)-h(x,t))^{q}}\rbrace^{1/q} \sim
l^{\alpha_{q}}$. Different exponents for $\alpha_{q}$: $\alpha_{2}
\simeq 0.6$, $\alpha_{4} \simeq 0.3$, and $\alpha_{6} \simeq 0.2$
were obtained, which confirm the existence of multiscaling in the
experiment.

\begin{figure}
\begin{center}  \epsfig{file=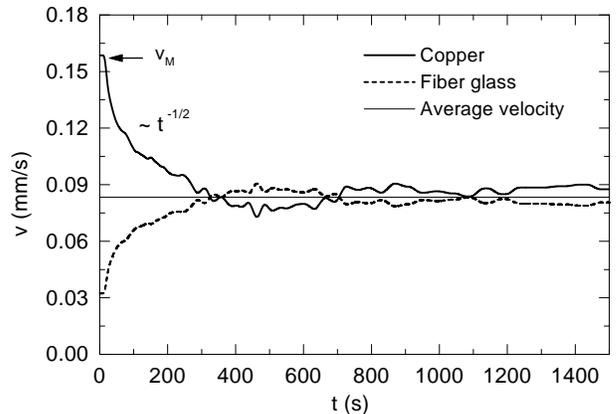,width=8cm}
\caption{Average interface velocity over
copper tracks and over fiberglass, for $b=0.36$ mm and $v=0.08$ mm/s.}
\label{Fig:Cu-fiber}
\end{center}
\end{figure}

\begin{figure}
\begin{center}  \epsfig{file=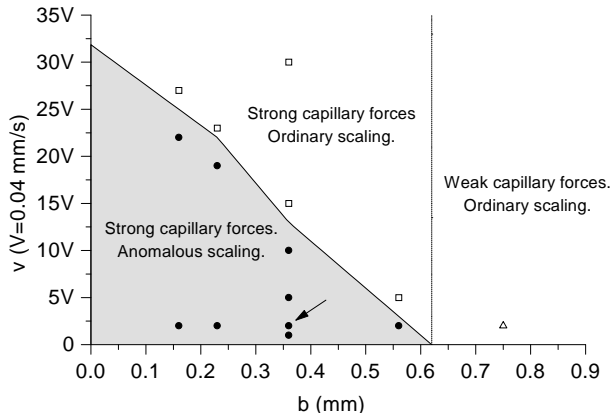,width=8cm}
\caption{Phase diagram $v(b)$  indicating the regions where the anomalous
scaling is observable. The symbols represent the different regions explored
experimentally, and the arrow indicates the parameters used in the paper.}
\label{Fig:diagram}
\end{center}
\end{figure}

To understand the origin of anomalous scaling, we should look at
the interfacial
dynamics in greater detail. An initially flat oil--air interface, moving at a
nominal average velocity $v$, experiences local accelerations at those points
that touch a copper track for the first time. The local velocity of these
points jumps to a maximum in a characteristic time of 1 s, and then relaxes as
$t^{-1/2}$ to $v$. The maximum velocity reached in a
copper track increases with track width and decreases with gap
spacing.
As shown in Fig.\ \ref{Fig:Cu-fiber}, the average interface velocity
over copper tracks or over fiberglass channels
follows $v_{\pm}=v \pm (v_M-v)t^{-1/2}$ respectively, where $v_M$ is
the maximum of the average velocity over
tracks. This functionality varies slightly with gap spacing due to the
increasing correlation between neighboring tracks. It does not apply for $b
\gtrsim 0.6$ mm.
This behavior of different velocities $v_{+}$ and $v_{-}$
is directly correlated with the shift of
the power spectra, indicative of anomalous scaling: the shift is maximum
at very short times and disappears close to saturation, similarly to the
difference between the average velocities shown in Fig.\ \ref{Fig:Cu-fiber}.
When we inhibit the relaxation of the velocity over copper tracks
by either injecting at $v > v_{M}$ or using large $b$, the
anomalous scaling disappears. This scenario has been studied experimentally by
exploring different velocities and gap spacings.
Fig.\ \ref{Fig:diagram} shows a phase diagram
representing the regions where the anomalous scaling is present (grey region,
$\theta \neq 0$) or not (white regions, $\theta=0$). The solid line represents
the function $v_{M}(b)$, and the vertical line the limit between strong and
weak capillary forces.

The critical exponents of the anomalous scaling depend on gap spacing and
velocity. We have seen that the anomalous exponent $\theta$ tends to
$\theta=0$ as we increase either the velocity or the gap spacing. This
exponent is particularly sensitive to variations of the gap spacing. For $b
\gtrsim 0.6$ mm, $\theta=0$ at any velocity. The experimental parameters
$b=0.36$ mm and $v=0.08$ mm/s, deep in the grey region of Fig.\
\ref{Fig:diagram}, give the appropriate conditions to fully characterize the
anomalous scaling experimentally. Smaller gap spacings mean that the
correlation length in the $x$ direction cannot grow to scales larger than the
average track width, and larger gap spacings give too weak an anomalous
scaling because the $t^{-1/2}$ law is only obeyed on the widest tracks.

Finally, in order to reproduce the experimental behavior in a
phenomenological way, we have developed an interfacial equation of the
form:
\begin{equation}\label{model}
\partial _{t} h = \partial_{x}D(x)\partial_{x}h + v +
(v_{M}-v)t^{-1/2}\eta(x),
\end{equation}
where $h=h(x,t)$, and $\eta(x)$ is a dichotomous noise with values
$+1$ and $-1$ (columnar disorder). The diffusive coupling accounts for
the dynamics with $z=2$ found in experiments, and the other two terms
account for the behavior of $v_{\pm}$ observed experimentally. Although the
term $t^{-1/2}$ introduces some nonlocality in the model, this is essentially
different from non--local models \cite{non-local-models} introduced
recently to account for mass conservation.
The simplest version (constant $D$) reproduces the experimental
behavior at long length scales
($\alpha=1$, $\beta=0.5$, $z=2$), but not at short length
scales, since numerical simulations of the model
give $\beta^{\ast} = 0$ and $\alpha_{loc} = \alpha = 1$.
The reason is that the simulated interfaces do not display the large
fluctuations at the copper track edges observed in the
experimental fronts (Fig.\ \ref{Fig:interf}).
This can be imposed analitically
in Eq.\ (\ref{model}) through an
inhomogeneous diffusion coefficient $D(x)$. Numerically, we have introduced
this effect by spatially averaging the interface inside each track once each n
time steps. We thus recover interfaces that are morphologically analogous to
experimental interfaces. Now the exponents derived from the scaling
of the power spectra, shown in Fig.\ \ref{Fig:collapse}(c), and the
multiscaling exponents, reproduce the values determined experimentally.
In summary, Eq.\ (\ref{model}) enables
the relevant effects in the experiment to be both identified and calibrated.

In conclusion, we have presented the first experimental evidence of
anomalous scaling in the dynamic roughening of a fluid interface in a
Hele--Shaw cell with quenched disorder, and we have studied the
experimental conditions for the appearance of the observed anomaly.  The
exponents $\beta$, $\beta^{\ast}$, and $\alpha_{loc}$ have been measured
independently and their value has been refined by imposing the scaling
relations through a collapse of the power spectra.  Finally we have
introduced an interfacial equation (\ref{model}) that models the
capillary phenomena observed in the experiments, at a phenomenological
level.  The model reproduces both the morphology of the interfaces and
the values of the anomalous scaling exponents.

We are grateful to J.M. L\'opez for fruitful discussions. The research has
received financial support from the Direcci\'on General de Investigaci\'on
(MCT, Spain) under projects BFM2000-0628-C03-01 and -02. J. Soriano and J.J.
Ramasco have received financial support from the Spanish Ministry of Science
and the Spanish Ministry of Education respectively.

\end{multicols}


\begin{references}
\bibitem{anomaly} The anomaly also decreases as the persistence of the
disorder in the advancing direction is reduced, e.g. when tracks are
replaced by square islands {\protect \cite{Soriano-2002-I}}.
\end{references}
\end{document}